\documentclass[12pt,preprint,longnamesfirst]{aastex}
\usepackage{epstopdf}
\usepackage{amsmath}

\newcommand{\kms}{\ifmmode \mathrm{km~s^{-1}}\else km~s$^{-1}$\fi}
\newcommand{\lsun}{\ifmmode L_\sun\else $L_\sun$\fi}
\newcommand{\msun}{\ifmmode M_\sun\else $M_\sun$\fi}
\newcommand{\lya}{\ifmmode \mathrm{Ly\alpha~}\else Ly$\alpha$~\fi}

\shorttitle{Searching for $z$$\sim$7.7 \lya Emitters in the Cosmos
  Field with NEWFIRM}
\shortauthors{Krug, Veilleux}
\begin{document}

\title{Searching for $z$$\sim$7.7 \lya Emitters in the COSMOS Field
  with NEWFIRM}
\author{Hannah B. Krug\altaffilmark{1}, Sylvain
  Veilleux\altaffilmark{1}, Vithal Tilvi\altaffilmark{2,3}, Sangeeta Malhotra\altaffilmark{2}, James
  Rhoads\altaffilmark{2}, Pascale Hibon\altaffilmark{2,4}, Rob Swaters\altaffilmark{4}, Ron Probst\altaffilmark{5}, Arjun Dey\altaffilmark{5}, Mark Dickinson\altaffilmark{5},
  \& Buell T. Jannuzi\altaffilmark{5}}
\altaffiltext{1}{Department of Astronomy, University of Maryland,
  College Park, MD 20742, USA; E-mail: hkrug@astro.umd.edu,
  veilleux@astro.umd.edu}
\altaffiltext{2}{School of Earth and Space Exploration, Arizona State
  University, Tempe, AZ 85287, USA}
\altaffiltext{3}{Department of Physics \& Astronomy, Texas A\&M
  University, College Station, TX 77843, USA}
\altaffiltext{4}{Gemini Observatory, La Serena, Chile}
\altaffiltext{5}{NOAO, Tucson, AZ 85719, USA}

\begin{abstract}
The study of \lya emission in the high-redshift universe is a
useful probe of the epoch of reionization, as the \lya line should be attenuated by the intergalactic medium (IGM) at low to moderate
neutral hydrogen fractions.  Here we present the results of a deep
and wide imaging search for \lya emitters in the COSMOS field.
We have used two ultra-narrowband filters (filter width of $\sim$8
-9 \AA) on the NEWFIRM camera, installed on the Mayall 4m telescope at Kitt Peak National Observatory, in
order to isolate \lya emitters at $z=$ 7.7; such ultra-narrowband
imaging searches have proved to be excellent at detecting \lya
emitters.  We found 5$\sigma$ detections of four candidate \lya
emitters in a survey volume of 2.8 x 10$^4$ Mpc$^3$ (total
  survey area
$\sim$760 arcmin$^2$).  Each candidate
has a line flux greater than 8 x 10$^{-18}$ erg s$^{-1}$ cm$^{-2}$.
Using these results to construct a luminosity function and comparing
to previously established \lya luminosity functions at $z=$ 5.7
and $z=$ 6.5, we find no conclusive evidence for evolution of the
luminosity function between $z=$ 5.7 and $z=$ 7.7.  Statistical Monte
Carlo simulations suggest that half of these candidates are real
$z=$ 7.7 targets, and spectroscopic followup will be required to verify
the redshift of these candidates.  However, our results are
  consistent with no strong evolution in the neutral hydrogen fraction
of the IGM between $z=$ 5.7 and $z=$ 7.7, even if only one or two of the $z=$ 7.7 candidates are
spectroscopically confirmed.

\end{abstract}


%%%%%%%%%%%%%%%%%%%%%%%%%%%%%%%%%%%%%%%%%%%%%%%%%%%%%%%%

\section{INTRODUCTION} \label{intro}

Direct observations of distant galaxies remain the most
straightforward way to probe the fundamental nature of the
high-redshift universe.  Such observations can provide some
much-needed constraints on numerical simulations, which may provide better answers to the question of
how large scale structure forms and how star formation begins in dark
matter halos. Star formation in early galaxies is dependent upon
the mechanism by which gas cools; that mechanism itself is dependent
upon the ionization state and metal enrichment of that gas which are
not well constrained at high redshift.  The
\lya emission line is a very useful tool for the detection of
high-redshift galaxies, as the earliest stars in the universe should
ionize surrounding hydrogen gas, which will then recombine to produce
\lya emission (see \citealt{wil08}).  This \lya line can be probed
quite effectively at high redshifts via the use of narrowband
filters, which focus on regions with low sky background and that are
free of strong OH lines (e.g., \citealt{cuby07}).  High-redshift objects should
have essentially no flux blueward of rest-frame \lya and none blueward
of rest-frame 912 \AA; this is a result of the \lya forest effect at high
redshift, due to strong absorption by intervening clouds (e.g.,\citealt{bah65,gp65,lynds71,rees86,mesc96,schaye01}.  Such narrowband surveys have proved quite
successful so far (e.g.,
\citealt{chu98,hu99,r00,fynbo01,ouchi01,hu02,mr02,ouchi03,r03,hu04,mr04,r04,tanig05,kash06,shim06,nils07,ouchi08,fink09})
and have resulted in samples of galaxies over a range of redshifts, including the spectroscopic confirmation of a \lya
emitting galaxy at $z$ $=$ 6.96 \citep{iye06}.  Even when these
narrowband surveys do not successfully detect high-$z$ objects,
such null results can be used to constrain the \lya luminosity
function (e.g., \citealt{cuby07,wil08,sobral09}).

The early universe is expected to be metal poor, but metals have
been detected in the intergalactic medium (IGM) at $z$ $=$ 5.7
\citep{rpm06}, and so the IGM must have been enriched in metals by $z$
$\sim$ 6 at the latest, with recent results tentatively indicating a metallicity
downturn between $z$ $\sim$ 5.7 and $z$ $\sim$ 5 \citep{simcoe11}.
Additionally, the \lya line is sensitive to IGM obscuration at
neutral hydrogen fractions ranging from low to high (e.g., $10\% 
\lesssim x_{HI} < 100\%$; \citealt{haim02,sant04}), and thus observations of
\lya-emitting galaxies serve as a powerful probe of the reionization
history of the universe; the Gunn-Peterson test, for example, is only
useful when the neutral gas fraction is $<$1\% \citep{cuby07}.
Increasing the neutral hydrogen fraction in the IGM increases the
attenuation of \lya emission from those galaxies \citep{sobral09}.  As
this neutral fraction increases, the \lya luminosity function will
vary according to the amount of light being attenuated by the IGM.
Previous estimates of the redshift of the epoch of reionization using
constraints from studies of \lya emitters do not concur with
constraints derived from polarization observations of the CMB.  The
latter suggests that the redshift of reionization is $z_\mathrm{re}$
$=$ 10.5 $\pm$ 1.2, should reionization be an instantaneous process \citep{kom11}, whereas the former have indicated a
significantly later end to the epoch.  Constraints on the \lya luminosity function (LF) can assist in
determination of the redshift at which reionization has been
completed, owing to the resonant scattering of \lya photons in a
neutral IGM.  If the intrinsic number density of young galaxies remains constant
over redshift, then a significant decline in the observed \lya LF at a
given redshift could indicate a change in IGM phase.  On the
lower redshift end, Malhotra \& Rhoads (2004) found no significant evolution of
\lya LF between 5.7 $<$ $z$ $<$ 6.6; whereas at higher redshifts, an
evolution of \lya LF between 6.5 $<$ $z$ $<$ 7 is suggested based on
single detections \citep{iye06,ota08}.  Ouchi et al. (2008) found
little evolution between $z$ $\sim$ 3 and $z$ $\sim$ 6 in the observed
LFs, although they suggest a real evolution, with increase in intrinsic
\lya luminosity being canceled out by increase in IGM absorption.
Curtis-Lake et al. (2011) have recently identified \lya emitters at a
high rate over 6 $<$ $z$ $<$ 6.5 in UKIDSS.  Ono et al. (2011) have measured a decrease in \lya emission line
detection fraction over 6 $<$ $z$ $<$ 7, as have Schenker et
al. (2011) over 6 $<$ $z$ $<$ 8 and Pentericci et al. (2011; following
up work by Fontana et al. (2010)) over 6 $<$ $z$ $<$ 7; these studies
are UV-continuum-selected galaxies (i.e., Lyman break galaxies) rather
than selected via \lya emission line but share the same goal, and all conclude that the
neutral hydrogen fraction of the IGM is
increasing over those epochs.  The seven \lya candidates found by
Hibon et al. (2010) at $z$ $=$ 7.7 would indicate no evolution of the
\lya LF if they are found to be spectroscopically confirmed (although
soon to be published work suggests that at least five are not
confirmed; see Cl\'ement et al. (2011)); the four \lya
candidates found by Tilvi et al. (2010) could either indicate mild
or no evolution between 6.5 $<$ $z$ $<$ 7.7 depending on the number of
candidates which are confirmed.  Two Lyman break galaxies at
  $z\gtrsim$7.0 were recently spectroscopically confirmed by
Vanzella et al. (2011), their luminosities fairly consistent with that
of the well-known confirmed $z=6.96$ Iye et al. (2006) \lya emitter.  At present, small number
statistics severely affect our ability to draw definite
conclusions on the \lya luminosity function and properties of the
IGM.  It is therefore essential to expand the sample of high-redshift
candidates at epochs when the IGM became metal-enriched and reionized,
and thus shed light on the nature of the
early universe.

As redshift increases, galaxy sizes and luminosities decrease, and
cosmological dimming must also be taken into account.  Because of
this, detection of galaxies at $z$ $>$ 7 can be quite difficult
\citep{ferg04,bou06,capak11}.  Thus it is essential to have a large
survey volume in order to detect a sufficient number of high-$z$
objects.  This conclusion is bolstered by the biased nature of galaxy formation
and the non-uniformity of large-scale structure at high-$z$
\citep{steid99,m05,wang05,ouchi05,tilvi09}.  Presently, most studies
searching for high-$z$ galaxies have significant depth but
small area (e.g., \citealt{bou10,oesch10}) or have small volume but
high magnification by virtue of cluster lensing
\citep{rich07,rich08,stark07}, so all results are affected by cosmic variance.  There is a need for surveys
which probe both a deep and wide region in order to best constrain
global properties of high-$z$ galaxies; the present paper reports on a
survey that tries to fill that role.

In this paper, we present the results from a search for \lya emitting galaxies at $z$
$=$ 7.7 in the COSMOS field, utilizing custom-made ultra-narrowband
filters which are tuned to avoid the OH sky lines and thus reach
extremely low infrared sky backgrounds.  The organization of this paper is as
follows.  In Section \ref{obsdata}, we discuss our observations and
reduction of the data that we obtained, as well as photometric
calibration.  In Section \ref{selection}, we describe our method of
candidate \lya emitter selection and the basic properties of the
resulting candidate \lya emitters.  In Section \ref{contaminants}, we discuss
possible sources of contamination in our samples.  In Section \ref{mcsim}, we estimate the number
of \lya emitting galaxies that we should expect to find in our survey
through the use of a detailed Monte Carlo simulation.  In Section
\ref{lylf}, we present the \lya luminosity function derived from
our candidates and compare to previously derived \lya LFs.  Finally,
in Section \ref{conc}, we summarize our conclusions.  This work shares
authors, instrument, and technique with the work of Tilvi et
al. (2010), and can thus be viewed as part of a series with that paper.  Throughout this
work, we assume a standard flat $\Lambda$CDM cosmology with $\Omega_m$
$=$ 0.3, $\Omega_\Lambda$ $=$ 0.7, and $h$ $=$ 0.71, where $\Omega_m$,
$\Omega_\Lambda$, and $h$ are the matter density, dark energy density,
and Hubble parameter (in units of 100 km s$^{-1}$ Mpc$^{-1}$),
respectively.  All magnitudes listed are in the AB magnitude system.

%%%%%%%%%%%%%%%%%%%%%%%%%%%%%%%%%%%%%%%%%%%%%%%%%%%%%%%%%%

\section{OBSERVATIONS AND DATA} \label{obsdata}

\subsection{Observations with NEWFIRM}
Our observations were centered on the Cosmological Evolution Survey (COSMOS) field (RA 10:00:28.6,
Dec. +02:12:21.0), taking advantage of the large amount of ancillary data
available on this
field\footnote{\texttt{http://cosmos.astro.caltech.edu}}.  We used the NOAO Extremely Wide-Field Infrared Mosaic (NEWFIRM) camera
on the Mayall 4m telescope at Kitt Peak National Observatory (KPNO)
on three different observing runs over the course of two years (2008
February 28 - March 14; 2009 January 29 - February 1; 2009 February 17
- March 2) for a total of roughly 100 hours over 32 nights.  Average
seeing over the course of these exposures was $\sim$1.2''.

\begin{figure}
\begin{center}
\plotone{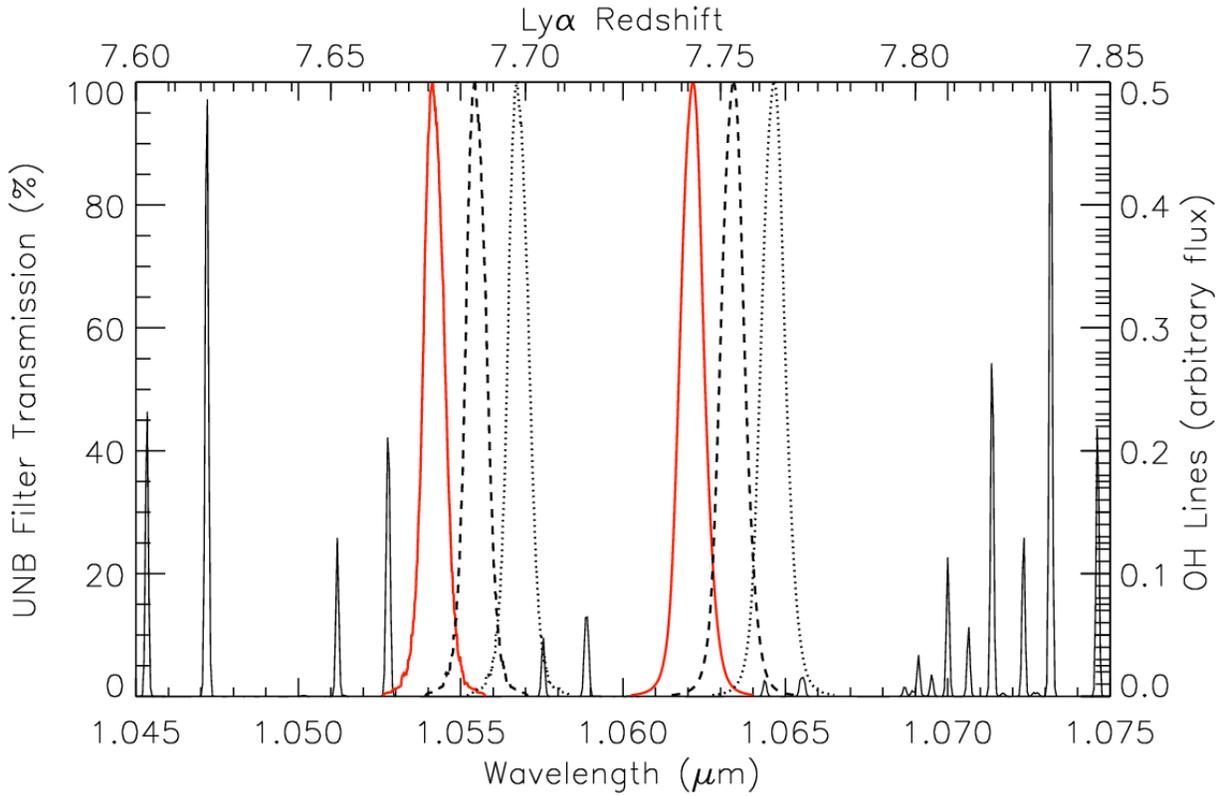}
\caption{NEWFIRM ultra-narrowband filter profiles, centered at 1.056
  and 1.063 $\mu$m.  The solid red, dashed, and dotted lines show the
  range of profile from center (solid red) to $\sim$ 40\% (dashed) and 80\%
  center-to-corner distance (dotted) of the field of view.  The
  Rousselot et al. (2000) OH sky spectrum is also plotted.}
\label{filters}
\end{center}
\end{figure}

The NEWFIRM camera, sensitive to 1-2.4 $\mu$m wavelengths, is a wide-field imager consisting of
four mosaiced 2048 x 2048 pixel ALADDIN InSb arrays, 0.4'' pixel$^{-1}$, for a
cumulative field of view of 27.6' x 27.6' (cumulative area $\sim$760
arcmin$^2$).  Our observing time was split evenly between two University of Maryland
custom-made ultra-narrowband (UNB) filters (R$\sim$1000) centered at 1.056 and 1.063
$\mu$m, with FWHM of 7.4 and 8.1 \AA, respectively.  These UNB filters were designed in order to isolate \lya
emitters at $z$~$\sim$~7.7, while simultaneously avoiding the bright OH
lines near these wavelengths (Figure~\ref{filters} shows UNB filter
profiles).  The transmitted wavelength varies across the field according to roughly $m\lambda$ $=$
$\mu d\cos\theta$ (here the order, $m$, index of refraction, $\mu$, and
thickness of the filter, $d$, are constant; $\theta$ is the angle of
incidence of light onto the detector); the path length increases as
one gets further off-axis and the wavelength of transmitted light is
correspondingly shorter.  This effect is more apparent 
for a UNB filter in a wide field imager, since the bandpass is narrow
and the range of angle of incidence is large (see Figure~\ref{filters}).   
Atmospheric absorption (primarily due to O$_2$ and H$_2$O in the IR) is irrelevant in these bandpasses.  Each 1200s exposure science frame was
taken using Fowler 8 sampling and we utilized random dithering within
a 45'' box after each exposure.  

Additional ancillary data in broadband filters were required for \lya
candidate selection.  We made use of publicly-available data from the COSMOS
archive\footnote{\texttt{http://irsa.ipac.caltech.edu/Missions/cosmos.html}}
for this purpose: $B$, $r$, and $i$ band data from Subaru and
$J$ band data from UKIRT.  

\subsection{Data Reduction}
The reduction of our data was done by the NEWFIRM Science Pipeline (see
Dickinson and Valdes 2009; Swaters et al. 2011).  Frames with seeing of $>$ 1.5'' were rejected
outright.  The pipeline flagged pixels affected by detector
blemishes, saturation, and persistence, and then subtracted the dark current,
linearized the data, and applied the dome flat. Image gradients remaining
after the flat fielding were subtracted out. The astrometric solution was
determined from 2MASS stars in the field; these same stars were also used to
determine an initial photometric calibration of the data. In the pipeline,
the sky was subtracted in a two-pass approach. First, the sky was subtracted
using a running-median window, and then the data were combined by taking the
median over all the exposures. This so-called harsh stack was then used to
identify and mask sources. The mask was then applied to the original images,
and the sky-subtraction was repeated. Cosmic ray hits and other transient
phenomena were detected by comparing individual images against the first-pass
stack, and outliers were flagged.

Starting with these pipeline products, custom IDL4 scripts designed by Krug \& Swaters were used to eliminate artifacts such
as OH rings and striping due to data readout from the science frames.
For OH ring
elimination, pixel values were separated into 1000 radial bins across
one NEWFIRM chip at a time.  Owing to the unevenness of the NEWFIRM chip gaps, the pixel center of the OH ring
varied from chip to chip; the ring centers were determined for each chip through a
combination of visual inspection and fitting rings to the values at
each pixel.  Data within the
radial bins were then median smoothed, and these smoothed bins were then subtracted from the
original image data.  Stripe removal was also performed on individual NEWFIRM chips at
one time.  Each chip was divided into horizontal or vertical strips, depending on data
readout direction.  Pixels across each strip were summed and averaged, and the original data
pixels in each strip were subtracted by the average pixel value in that strip. A comparison of
stacks made from frames before and after ring and stripe removal showed that the
signal-to-noise ratio (SNR) increased by 1.2 times on average across the frames, with SNR increasing by
as much as 4-7 times in regions where the OH rings were strong prior to algorithm
implementation.  The SNR in the most extreme rings, however (namely
those areas as seen in Figure~\ref{filters} where the OH lines cross
the outer edge filters, particularly the strongest OH ring in the
1.063 $\mu$m filter), was not quite high enough for candidate
detection (see Section~\ref{selection}).

Following removal of artifacts and sky subtraction, the world coordinate
system of each individual frame was assigned using the IRAF task $nfwcs$, which matches as many
sources as possible to those of the 2MASS point source catalog, searching within a
450'' radius. The $mscimage$ task was used to reproject all four chips onto a single image with
the same pixel grid. Finally, images were combined into two yearly
stacks (weighted by seeing) as well as one full
data stack for each filter via $imcombine$; these were median combined using noise values from
$mscstat$ and zeropoint magnitudes as calculated based on the 2MASS catalog during the prior
$nfwcs$ task. This resulted in two full UNB stacks - one each for the 1.056 and 1.063
$\mu$m filters - as well as stacks from year one and year two for each filter (in order to
properly check for transients). Broadband data mosaics supplied by the COSMOS team were
already reduced. 

\subsection{Photometric Calibration} \label{photocal}
Photometric calibration was performed by making use of the 2MASS point
source catalog.  SExtractor (SE; \cite{ba96}) was run on each NEWFIRM UNB stack using a
magnitude zero point set to zero.  The resulting catalogs were matched with
the 2MASS point source catalog; in order to avoid saturated sources or
objects that are too faint, all stars with 2MASS $J$ band magnitudes (AB) of
less than 13.0 or greater than 16.5 were discarded.  Plots were made
comparing the SE UNB magnitudes with 2MASS $J$ magnitudes in
order to determine an appropriate zeropoint.  Before SE could be run with the proper magnitude
zeropoint, it was necessary to determine a color correction between
the 2MASS $J$ magnitude and the NEWFIRM UNB magnitudes, as they do not
share the same central wavelength.  2MASS $J$, $H$, and $K$ band
magnitudes for overlapping stars were converted to flux and used to
calculate the flux expected at 1.056 and 1.063 $\mu$m, the central
wavelengths of the NEWFIRM UNB filters.  The difference between 2MASS $J$ flux and extrapolated NEWFIRM UNB
flux corresponds to a magnitude difference of $\sim$+0.1 mag in $J$.  This
correction was then applied to the previously determined magnitude
zeropoint.    

\subsection{Limiting Magnitude} \label{limmag}
\begin{figure}
\begin{center}
\plottwo{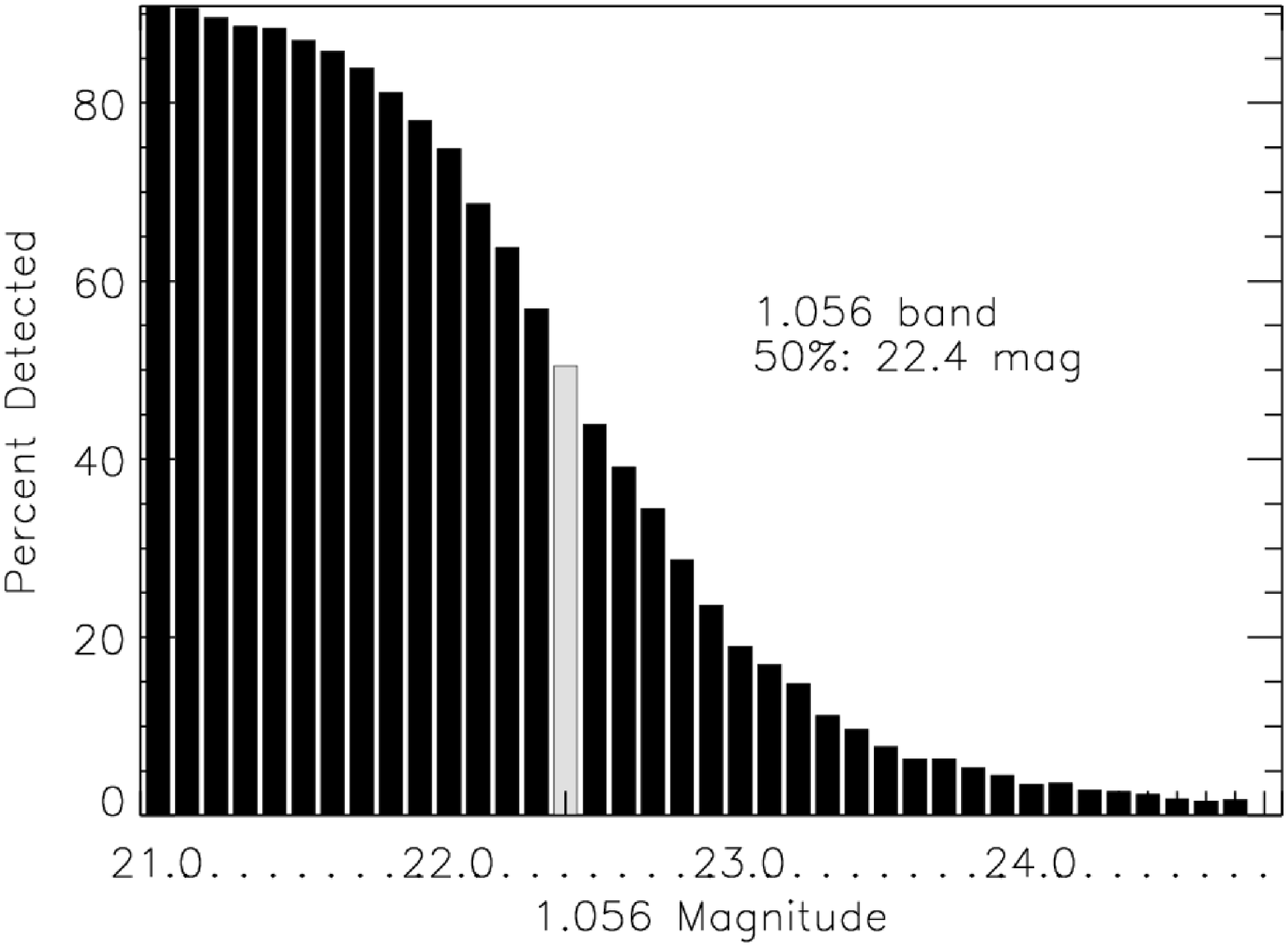}{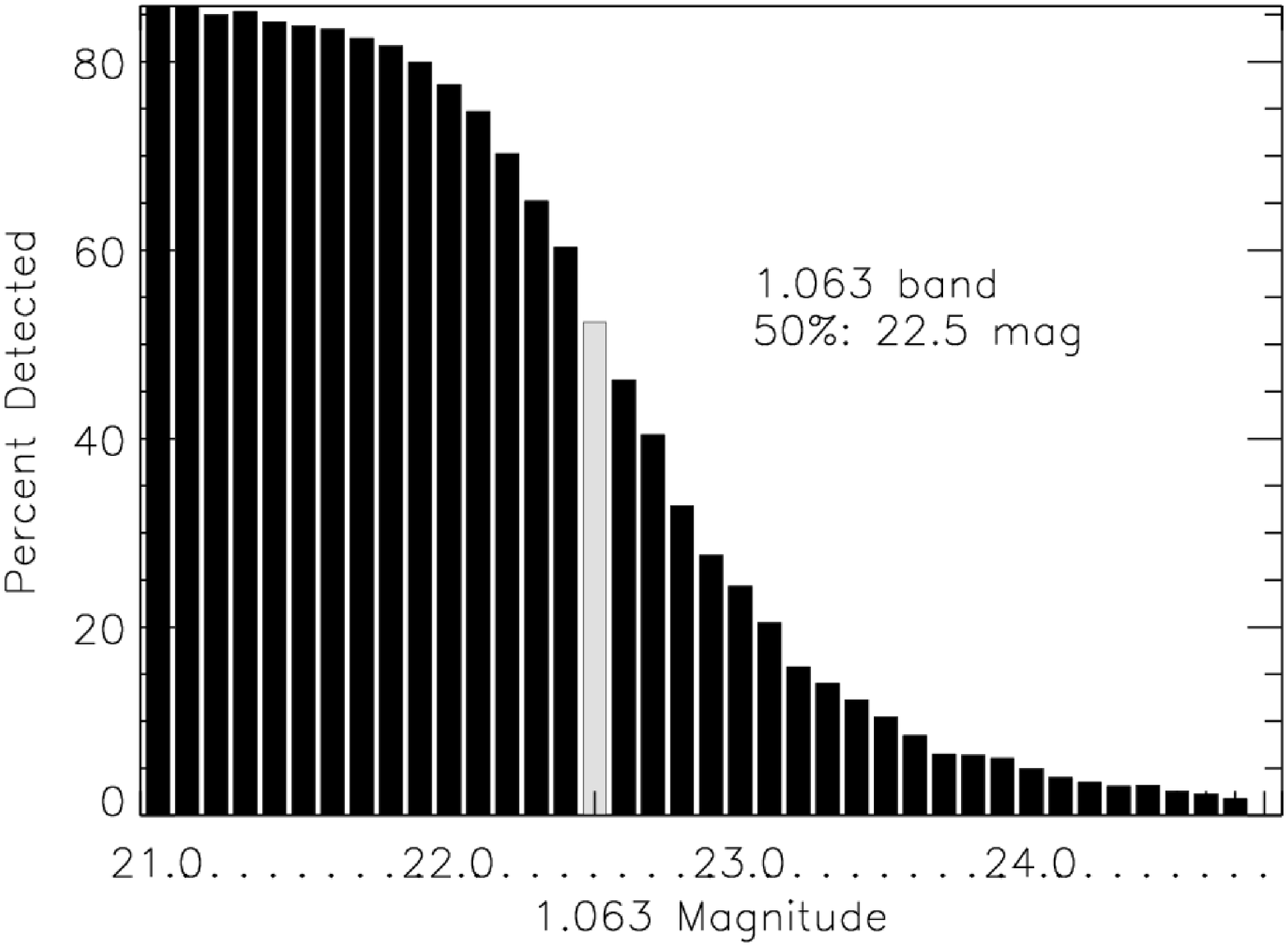}
\caption{SExtractor detection percentages for randomly generated
  false sources in 0.1 magnitude bins in a given UNB band.  The light gray
  bar marks 50\% completeness.  That value is used as the limiting
  magnitude (AB) for our survey.  Our 1.063 $\mu$m band data is 0.1 mag
  deeper than our 1.056 $\mu$m band data.}
\label{completeness}
\end{center}
\end{figure}

In order to determine which of our detections can be
considered real, we must obtain some estimate of the limiting
magnitude of our science frames.  We have chosen to define our
limiting magnitude as the 50\% completeness limit in our frames.  To determine this, we inject artificial
sources into our frames.  These sources are randomly distributed
throughout each frame, although care is taken to avoid locations within
4'' of any existing bright star.  Two hundred artificial point
sources are
generated in each of 0.1 magnitude bins, ranging between 20 and 24 mag
for UNB frames, 21 and 25 for $J$ band, and 24 and 28 mag for the
broadband optical frames.  Once these
sources were generated, SE was run and the resultant SE catalog was
matched with the catalog of artificial sources to determine a recovery
fraction.  This procedure was then iterated 25 times for each band.
Once all iterations were completed, histograms of recovery fraction of
artificial sources were plotted for each band in order to determine the
magnitude at which 50\% of all artificial sources were detected by
SE  (Figure \ref{completeness}~for UNB recovery fraction
histograms).  We define the limiting magnitude for each band as the 50\%
completeness limit for that filter, as this is the magnitude brighter
than which sources can be reliably detected in these fields via
our methods.  Limiting magnitudes for the two NEWFIRM UNB bands were
determined to be 22.4 and 22.5 (AB) for 1.056 $\mu$m and 1.063 $\mu$m,
respectively.  Shifts in central wavelength of the UNB filters with
distance from the field center were not found to significantly affect
these limiting magnitudes.  For the broadband data, limiting magnitudes were found
to be 27.3, 26.6, 26.3, and 24.0 for $B$, $r$, $i$, and $J$, respectively.

\section{CANDIDATE \lya EMITTERS} \label{selection}

\subsection{Selection Criteria}
To select potential candidate \lya emitters, we initially ran SE on
the NEWFIRM UNB frames individually, with minimum detection area of 5
pixels and detection threshold of 1.2 sigmas.  We then used dual-image mode in
SE, which takes the coordinates of objects detected in the NEWFIRM UNB frames
but measures fluxes at those coordinates in the UKIRT $J$ band image.
Thus we obtained $J$ band fluxes at the exact locations of NEWFIRM UNB
targets.  For all of our dual-image mode detections, we made use of the
SExtractor auto aperture size scaling.  We also needed to
obtain optical fluxes at those coordinates, so in order to do so, we
created a chi-squared image of a combination of $B$, $r$, and $i$
band data from
Subaru using the program $Swarp$ \citep{bertin02}.  The chi-squared
image is created by deriving the distribution of the pixels that
are dominated by object flux rather than those dominated by sky noise
\citep{szalay99}.  This is an optimal combination of images which
enhances real objects in the resultant frame rather than a simple
stack.  We then ran SE
in dual-image mode on this combined optical image in the same vein as
the $J$ images.  We also ran dual-image mode on the two NEWFIRM UNB
frames together - first using the coordinates for 1.056 $\mu$m
detections but getting 1.063 $\mu$m flux, and vice versa - as any
potential candidate \lya emitter should appear in one of the UNB
frames but not both.  We likely miss highly extended \lya
  blobs with this technique, especially following sky subtraction, but
  such highly extended sources would be difficult to isolate in any case.

Once the SE catalogs were created, we ran a custom python script to
comb through those catalogs and identify potential candidates.  There
were five main initial criteria that had to be met in order for an
object to be considered a candidate ($J$ data refers to the UKIRT band,
centered at 1.2$\mu$m and outside the UNB filter range):

\begin{enumerate}
\item UNB signal-to-noise ratio (SNR) of at least 5.
\item UNB excess $f_{\nu}$(UNB) - $f_{\nu}$($J$) of at least 3$\sigma$.
\item Flux ratio $f_{\nu}$(UNB) / $f_{\nu}$($J$) of at least 2.
\item UNB SNR in the other UNB filter of no more than 2
  (e.g. if detected in 1.056 $\mu$m, must not have a 1.063 $\mu$m SNR
  of more than 2).
\item Optical SNR in the chi-squared broadband image of no more than
  2.
\end{enumerate}

These selection criteria have been utilized in lower redshift ($z$ =
4.5, 5.7) searches for \lya emitters and have detected those emitters
at a 70-80\% success rate after spectroscopic confirmation, and we are
confident that these criteria will translate to higher redshift, as
the fundamental physics of the \lya forest should not change, and similar criteria have been used in a successful spectroscopic search at
$z$ $=$ 6.96 (see,
e.g., \citealt{rm01,r03,daws04,iye06,daws07,wang09}).  These criteria
are also the same as those used in the work of Tilvi et al. (2010).  The first three criteria are used
to isolate emission line sources.  The remaining two are used to
eliminate as many low-redshift sources as possible; as detailed in
Section \ref{intro}, \lya emitters should have no flux blueward
of their \lya emission line, and the \lya line should be narrow enough
that it is only detected in one NEWFIRM UNB filter and not both.

Following execution of the selection script, we were left with 65 potential candidates out of 31254 initial detections for the 1.056
$\mu$m band, and 110 potential candidates out of 32382 initial
detections for the 1.063 $\mu$m band.  We then matched our target
lists to the existing COSMOS source catalogs (obtained from Peter
Capak of the COSMOS team; personal communication).  All targets which
met our selection criteria but which were also in the COSMOS catalog
with a measured photometric redshift were set aside (Section
\ref{interlopers}~for further information on these low-redshift
interlopers).  Once the low-$z$ interlopers were removed, target lists were then
narrowed down further through a variety of methods: eliminating all
targets on the edges of the chip which lie in or outside the main
OH-line ring; eliminating all targets that lie within two arcseconds of
the chip gap (5 pixels); eliminating all targets within two arcseconds from a very
bright star.  Roughly 90\% of the initial non-interloper candidates were removed
this way.  Basic visual inspection was then performed as a sanity
check.  Our targets were also compared with the yearly stacks for
each filter (both via SE and visual inspection) in order to ensure
that these targets were not transients (Section
\ref{contaminants}).  As a final sanity check, we did not consider any candidates with
magnitudes fainter than the limiting magnitudes for each
filter (Section \ref{limmag}).

\subsection{Results}

\begin{figure}
\begin{center}
\plotone{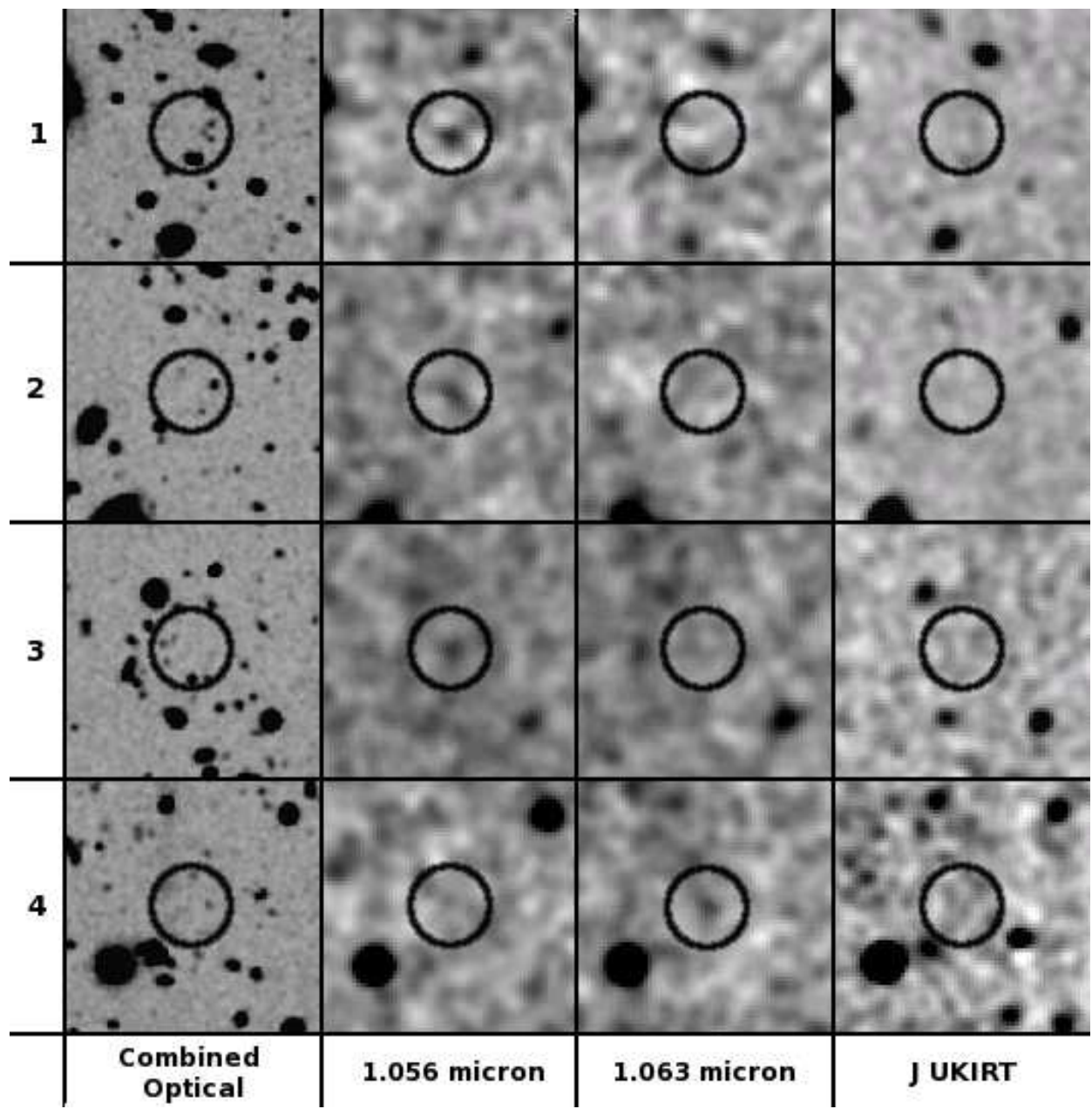}
\caption{Image cutouts for our four \lya emitter
  candidates.  UNB cutout images have been Gaussian smoothed according to the
  average seeing of the 1.056 and 1.063 $\mu$m data.  Cutouts are 50'' on each side, and the circles are 16'' in
  diameter (corresponding to $\sim$400 kpc at $z = 7.7$).  Each row represents one candidate.  The optical column
  shows a weighted chi-squared combination of $B$, $r$, and $i$ band
  images from Subaru.  The middle two columns show our UNB NEWFIRM
  data.  The $J$ band column represents data from UKIRT.  Candidates
\#1-3 are detected in 1.056 $\mu$m but not in any other band;
  candidate \#4 is only detected in 1.063 $\mu$m.  All other objects
  present only in one UNB band are either transients (detected in only
  one yearly UNB stack) or fail to meet the \lya selection
  criteria.}
\label{cutouts_high}
\end{center}
\end{figure}

After all these tests were completed, we were left with a total of 4
candidates brighter than the 50\% completeness limit - 3 candidates in
1.056 $\mu$m, 1 in 1.063 $\mu$m.  Three of these candidates lie at the
survey line flux limit.  The coordinates and basic
properties of these four luminous candidates - AB magnitude, line flux,
and luminosity
- are listed in Table \ref{cands_high}.  Line flux was calculated from the
SExtractor magnitude using:

\begin{equation}
\mathrm{F}~=~10^{-0.4*(\mathrm{mag}_\mathrm{AB} + 48.60)}~\dfrac{c}{\lambda^2}~
  \mathrm{W}~~\mathrm{ergs^{-1} cm^{-2} s^{-1}},
\end{equation}\\
where mag$_\mathrm{AB}$ is the magnitude from the isophotal fit as reported by SExtractor, $c$ is the speed of light, $\lambda$ is the central wavelength
of the given filter, and W is the filter width (6.95 $\times 10^{-4}$
$\mu$m for the 1.056 $\mu$m filter, 7.49 $\times 10^{-4}$ $\mu$m for
1.063).  Whereas we used the SE auto fit for initial dual-image mode detection, we use
isophotal fit for flux calculation to avoid losing signal and gaining
noise.  We can also
estimate the star formation rates for these objects, assuming that
that there is little attenuation of the \lya line by the neutral IGM
and that the dust content along the line of sight is low.  We use the
following prescription from Ota et al. (2010), which uses the
Kennicutt law \citep{kenn98} and
assumes case B recombination:
\begin{equation}
\mathrm{SFR}(\lya)~=~
9.1~\times~10^{-43}~\mathrm{L}(\lya)~\mathrm{M_\odot~yr^{-1}}.
\end{equation}
Using this calculation, we find that star formation rates for our four
\lya candidates range from 5 to 7.6 M$_\odot$ yr$^{-1}$.  Postage-stamp images of these
candidates are shown in Figure \ref{cutouts_high}.  Note that each
candidate is visible in one particular NEWFIRM UNB stamp but is not
visible in the other NEWFIRM UNB stamp.  Additionally, no candidate is
visible either in the chi-squared broadband optical stamp nor the
UKIRT $J$ band stamp.  As stated in Section \ref{selection}, in order
to eliminate most foreground galaxies, it is required that none of the
candidates are detected in the broadband image; non-detection in the
$J$ band image is simply due to the faint continuum of the \lya
candidates.  As can be seen in Figure \ref{cutouts_high}, the PSF is
considerably better in the Subaru band stack than in our UNB data.  To
try to determine whether this could affect our
candidate selection, we have attempted a modified aperture correction
using a comparison between total SE auto flux in the Subaru field and
flux derived from isophotal fits; this correction allows us to account
for faint sources
in the optical image.  Following this correction, we
re-ran our selection criteria.  Three of our four high-$z$ Ly$\alpha$
candidates still passed the selection criteria, but candidate \#3 failed
selection following this aperture correction (optical SNR $>$ 2).  We
still include candidate \#3 in the following analysis but flag it in
Table \ref{cands_high}~as more uncertain than the other candidates.

\subsection{\lya Equivalent Widths} \label{eqw}

As we have a sample of four strong candidate \lya emitters, it is
worthwhile to compare the equivalent widths (EWs) of their \lya emission
lines to those already noted in the literature.  Several published studies have spectroscopically identified \lya emitters with rest-frame
equivalent widths of EW$_{\mathrm{rest}}$ $>$ 240\AA~at $z$ $=$ 4.5
and 5.7, significantly
higher than predicted by theoretical simulations of star-forming
galaxies \citep{mr02,shim06,daws07,gron07,ouchi08}.

The calculation for rest-frame EW for our \lya emitters makes use of
the fluxes measured in both the NEWFIRM UNB filters and the broadband UKIRT
$J$ data, as the \lya line does not appear in the $J$ band (again, the
$J$ band data is centered at 1.2$\mu$m, with no overlap with our UNB filters).  However, none of our candidates were detected in the $J$ band, and thus we must use the $J$
band limiting magnitude (Section \ref{limmag}) to calculate an
upper limit on $J$ band flux.  In this case, given the limiting
magnitude of 24.0 in the UKIRT $J$ band, the upper limit on the $J$ band
continuum flux $f_{\lambda,J} = $
1.9 $\times$ 10$^{-19}$ erg s$^{-1}$ cm$^{-2}$ \AA$^{-1}$.  This value
is used to calculate a lower limit on the rest-frame \lya EW for
our candidates:

\begin{equation}
\mathrm{EW}_{\mathrm{rest}} = \dfrac{f_{\mathrm{UNB}}}{f_{\lambda,J}}
\times \dfrac{1}{1+z},
\end{equation}\\
where $f_{\mathrm{UNB}}$ is the UNB line flux in erg s$^{-1}$
cm$^{-2}$.  This calculation assumes that the \lya line falls entirely
within the transmission profile of
our UNB filters.  In
reality, the UNB filters may not enclose all of the \lya emission.
Moreover, we are using upper limits on the $J$ band fluxes.  The EWs
measured in this way should thus be viewed as 1-$\sigma$ lower limits.  This
calculation also assumes an exact redshift of $z$ $=$ 7.7.  

The resultant lower limits on the rest-frame EW (\lya) are 7.32, 5.17, 4.89, and 4.84
\AA~for candidates 1-4, respectively.  Thus our lower limit on EW for our candidates is
EW$_{\mathrm{rest}}$ $\gtrsim$ 4.8 \AA, much more consistent with
theoretical predictions than the numbers quoted above for lower
redshift surveys.  Our lower limit EW is smaller than that
of Hibon et al. (2010) by several angstroms, primarily owing to the
difference in bandwidths of UNB filters used by these two surveys.
This line width is also larger by $\sim$2 \AA~than the results of the
Tilvi et al. (2010) survey,
although this difference can be accounted for by the increase in depth
of our $J$ band data (limiting magnitude of 24.0 for our data versus
23.5 for their data).  As future surveys obtain deeper $J$ band data, we
should be able to better constrain the lower limits of \lya EWs for these
emitters. 

\section{POSSIBLE SAMPLE CONTAMINATION} \label{contaminants}
There are several possible sources of contamination in our sample of
candidate \lya emitters.  These include such
real sources as foreground emitters, transients, and cool L
\& T dwarfs, as well as false sources such as detector noise spikes
and general false detections.  We discuss each class of contaminants below.  

\subsection{Foreground Emission Line Sources} \label{interlopers}
There are three main species of foreground emission line objects which
are most likely to contaminate our sample, as each species should have a
strong emission line that falls within our UNB filter window and,
assuming faint continuum emission, negligible flux in nearby blue- and redward bands.  These species are
H$\alpha$ $\lambda$6563 emitters at $z$ = 0.62, [\ion{O}{3}]
$\lambda$5007 emitters at $z$ = 1.12, and [\ion{O}{2}]
$\lambda$3727 emitters at $z$ = 1.85.  Given the extensive
amount of ancillary data available for the COSMOS field, we were able
to first check our catalog with the main COSMOS catalog and eliminate
any low-$z$ interlopers from our \lya candidate list, as mentioned in
Section \ref{selection}.  A total of 3 interlopers fulfilled the five
main \lya emitter selection criteria (Section \ref{selection}), all in the 1.063
$\mu$m filter -- one brighter than the limiting magnitude of the field
-- which were removed from the sample of \lya candidates.  These faint
objects, however, are not at the exact redshifts listed above: the
COSMOS photometric redshifts are $z$ $\sim$ 1.5 $\pm$ 0.2 (1$\sigma$;
possibly corresponding to redshifted H$\gamma$)
, $z$ $\sim$ 1.7 $\pm$ 0.2 (possibly [OII] $\lambda$3727), $z$ $\sim$
2.5 $\pm$ 0.2 (possibly broad Mg II $\lambda$2798), but the errors
may be even larger.  It is possible that there remain some foreground
emission line objects in our \lya candidate sample that do not have
tabulated photometric redshifts.  Our survey benefits greatly from the
multi-wavelength coverage and accurate photometric redshifts of the
COSMOS data catalog; the presence in our sample of interlopers with photo-$z$s
different from the three expected species listed above may imply that
surveys in other fields without such comprehensive ancillary data
coverage suffer from greater low-$z$ contamination than estimated.

To estimate the number of additional possible emission line source interlopers
that could remain among our candidates, we must estimate the minimum
equivalent width required for these emission lines to contaminate our
sample.  We must also use the depth of our UNB image in order to
calculate the minimum luminosities of these emission lines.  As
described in Section \ref{limmag}, the limiting magnitude of our 1.056
$\mu$m stack is 22.4, which is equivalent to a 50\% completeness limit
in flux of 7.4 $\times$ 10$^{-18}$ erg s$^{-1}$ cm$^{-2}$.  Given the
redshifts of the aforementioned emission lines and using $CosmoCalc$\footnote{\texttt{http://www.astro.ucla.edu/$\sim$wright/CosmoCalc.html}}
to calculate luminosity distances, we find that the minimum
luminosities required to detect these emitters are 1.15 $\times$
10$^{40}$ erg s$^{-1}$, 4.96 $\times$ 10$^{40}$ erg s$^{-1}$, and 1.72
$\times$ 10$^{41}$ erg s$^{-1}$ for H$\alpha$, [\ion{O}{3}], and
[\ion{O}{2}] respectively.  To calculate the necessary minimum
equivalent width (observer-frame), we use the prescription from Rhoads \&
Malhotra (2001):

\begin{equation}
\mathrm{EW}_{\mathrm{min}} \equiv
      \left[ \dfrac{f_{\mathrm{nb}}}{f_{\mathrm{bb}}} - 1 \right] \Delta
      \lambda_{\mathrm{nb}} =
      \left[ \dfrac{5\sigma_{\mathrm{nb}}}{2\sigma_{\mathrm{bb}}} - 1 \right]
      \Delta \lambda_{\mathrm{nb}},
\end{equation}\\
where $f_{\mathrm{nb}}$ and $f_{\mathrm{bb}}$ refer to the flux
densities of the UNB and the chi-squared optical image, respectively,
$\sigma_{\mathrm{nb}}$ and $\sigma_{\mathrm{bb}}$ refer to the flux
measurement uncertainties in the two frames, and $\Delta
\lambda_{\mathrm{nb}}$ is the width of the UNB filter.  For these
calculations, we have simply used the 1.056 $\mu$m UNB filter.  It is safe to
assume that the continuum contribution to the overall flux in the UNB
filter is negligible, since the overall width (effective FWHM) of the
transmission profile of the 1.056 $\mu$m filter is only 9 \AA~(6.95 \AA).  For our filters, $\sigma_{\mathrm{nb}}$ =
1.06 $\times$ 10$^{-29}$ erg cm$^{-2}$ s$^{-1}$ Hz$^{-1}$ and $\sigma_{\mathrm{bb}}$ = 2.1 $\times$ 10$^{-31}$ erg
cm$^{-2}$ s$^{-1}$ Hz$^{-1}$.  This results in a minimum equivalent
width for the foreground emission line contaminants of EW$_{\mathrm{min}}
\gtrsim$ 870\AA.  

Although the distribution of equivalent widths of
these three species of emitters has not been probed at the high redshifts
of our observations, we can scale the published results under the
assumption that the luminosity functions of these species have not
evolved significantly between the relevant redshifts.  One particular recent
study has probed emission line sources of H$\alpha$ at $z$ = 0.27, [\ion{O}{3}] at $z$ = 0.51,
and [\ion{O}{2}] at $z$ = 1.0 in the GOODS-South field
\citep{str09}.  The minimum EW calculated above was scaled
appropriately in each instance to the EW one would expect at the
redshifts probed by the Straughn et al. (2009) survey.  Using that value and
the required minimum luminosity listed above, we determined how many
emission line objects we should expect in our field, after scaling
appropriately by the ratio of survey volumes.  We find that we expect
less than one additional interloper for each species (0.1 each for
H$\alpha$ and [\ion{O}{2}], 0.3 for [\ion{O}{3}]).  If we relax the
minimum equivalent width criterion to 500 \AA~prior to scaling EWs to
what would be expected at $z$ $=$ 7.7, the expectation increases to 1 additional interloper each for
H$\alpha$ and [\ion{O}{2}], and 2 additional interlopers for [\ion{O}{3}]. 

\subsection{Other Possible Contaminants}

The utilization of individual yearly stacks, as well as nightly
stacks, which were produced by the data reduction pipeline, allowed us
to eliminate both possible detections due to noise spikes as well as
real sources such as transient objects.  We required during our
candidate selection process that potential candidate \lya emitters be
detected in all stacks as well as the overall stack, though the
detection requirements for individual nightly/yearly stacks were less
stringent.  This eliminates contamination by noise spikes from the detector, as such
noise spikes should not be present at the same coordinates across multiple nights over a span
of years.  Transients will also be eliminated by this yearly stack check, as
supernovae should only be visible for a few weeks, not years.  We
were able to eliminate upward of twenty contaminants via this
requirement.  

In order to determine whether any false detections were contaminating
our candidate list, we checked to see whether taking the negative of
our UNB stacks would result in a detection.  We used IRAF to multiply each UNB
stack by -1 and then ran SExtractor on the resultant negative images.
The SE negative stack catalogs were then run through the exact same selection process
as the positive stack catalogs, but no candidates were identified this
way.  We are thus confident that the probability of false detections
contaminating our \lya sample is insignificant.

The final source of potential contaminants consists of cool stars - namely, L and T dwarfs - passing our
selection process.  We use previously observed relationships between spectral
type and absolute magnitude \citep{tin03} to determine in what $J$ band
magnitude range L and T dwarfs fall, and then we use our
calculated $J$ band limiting magnitude to determine distance ranges at
which we should be able to see these objects.  We find that L dwarfs
could be detected between roughly 550 and 1700 pc, and T dwarfs
could be detected between roughly 200 and 750 pc.  These L and T
dwarfs are most likely found within a Galactic disk scale height of
350 pc, however \citep{ryan05} -- the number density drops
significantly above the scale height of the disk -- and thus knowing that
these dwarfs typically have a volume density of no more than $\sim$3
$\times$ 10$^{-3}$ pc$^{-3}$, we thus can conclude that we should
expect at most one L or T dwarf in our survey.  If we then take our
selection criteria into account, we can determine if any L and T
dwarfs would satisfy our narrowband excess criterion.  Tilvi et
al. (2010) used existing observed spectra of L and T dwarfs and
a calculation of expected flux through our NEWFIRM filters.  They
determined that the flux which would be transmitted through the NEWFIRM UNB filters,
in comparison to that which would be transmitted through the NEWFIRM $J$
filter, would not be sufficient to pass our narrowband excess
criterion.  Thus, we should expect that no L nor T dwarfs would pass
our selection criteria and contaminate our sample. 

\section{MONTE CARLO SIMULATIONS} \label{mcsim}

As discussed in Section \ref{contaminants}, we do not expect all of
our four candidate \lya emitters to be real, as it is possible that
further low-$z$ emission line source interlopers may pass our
selection criteria.  In order to determine how many of our \lya candidates we
should expect to be real, we run Monte Carlo statistical simulations using
previously known \lya luminosity functions at lower redshift.  In this
work, we make use of the Kashikawa et al. (2006) \lya luminosity function at
$z$ = 6.5 and assume that there has been no significant evolution
between $z$ = 6.5 and $z$ = 7.7.  These simulations also make use of
the fact that the NEWFIRM UNB filter transmission curves may not
encompass the full width of the expected \lya emission lines in these
objects, and thus our measurements may in fact be underestimating the
\lya line flux produced by our \lya candidates.  The transmission
curves of our two UNB filters were obtained from the NEWFIRM
website\footnote{\texttt{http://www.noao.edu/ets/newfirm/documents/1056\%20nm\%20data\%20pack.xls}}$^,$\footnote{\texttt{http://www.noao.edu/ets/newfirm/documents/1063\%20nm\%20data\%20pack.xls}}.  

To begin our Monte Carlo simulations, we utilized the aforementioned
Kashikawa et al. (2006) luminosity function to generate one million random
galaxies.  These galaxies were distributed with both a random
\lya luminosity in the range of 10$^{42}$ erg s$^{-1}$ to 1.5 $\times$
10$^{43}$ erg s$^{-1}$, as well as a random redshift in the range
probed by our filters.  In the case of the 1.056 $\mu$m band, this range is 7.66
$<$ $z$ $<$ 7.71; for the 1.063 $\mu$m band, this range is 7.72 $<$
$z$ $<$ 7.76; we chose these ranges to correspond to the full width
of the transmission profile at which the transmission falls to 5\%. 

From these luminosities and redshifts, we were then able to assign a
flux to each of our randomly generated galaxies.  We made use of an
  asymmetric \lya line flux profile based on spectra of $z$ $=$ 5.7
  \lya emitters taken by Rhoads et al. (2003).  We were able to then use the
  detailed NEWFIRM UNB filter transmission curve profiles to derive
  the flux that would pass through the filter via a convolution of
  filter profile and line flux according to $f_{\mathrm{trans}}$
   $=$ $\int f_\lambda T_\lambda d\lambda$, where $T_\lambda$ is the
    NEWFIRM UNB filter transmission curve and $f_\lambda$ is the flux
    density of the emission line based on the Rhoads et al. (2003) spectra.
    This takes the likely underestimation of \lya line flux into
    account.  Following this, we were able to convert the convolved
    line flux into a magnitude by the use of the formula: 

\begin{equation}
\mathrm{mag_{AB}} = -2.5 log_{10} \left(\dfrac{f_{\mathrm{trans}}}{f_0}\right),
\end{equation}\\
where 

\begin{equation}
f_0 = \dfrac{3.6 \mathrm{kJy} \times \mathrm{c}}{(1.06 \mu m)^2} \times
\int T_\lambda~d\lambda~\mathrm{erg s^{-1} cm^{-2}},
\end{equation}\\
and $c$ is the speed of light.  In order to ensure that all
instrumental effects were taken into account, the last step in this
process was to incorporate the detection fraction at each magnitude
bin (Section \ref{limmag}; Figure \ref{completeness}).  This
detection fraction was multiplied by the number of galaxies in each
magnitude bin before we converted those magnitudes to \lya
luminosities.  The number of detected galaxies in each \lya luminosity
bin was then used to estimate how many \lya emitters we should expect
from our survey.

\begin{figure}
\begin{center}
\plottwo{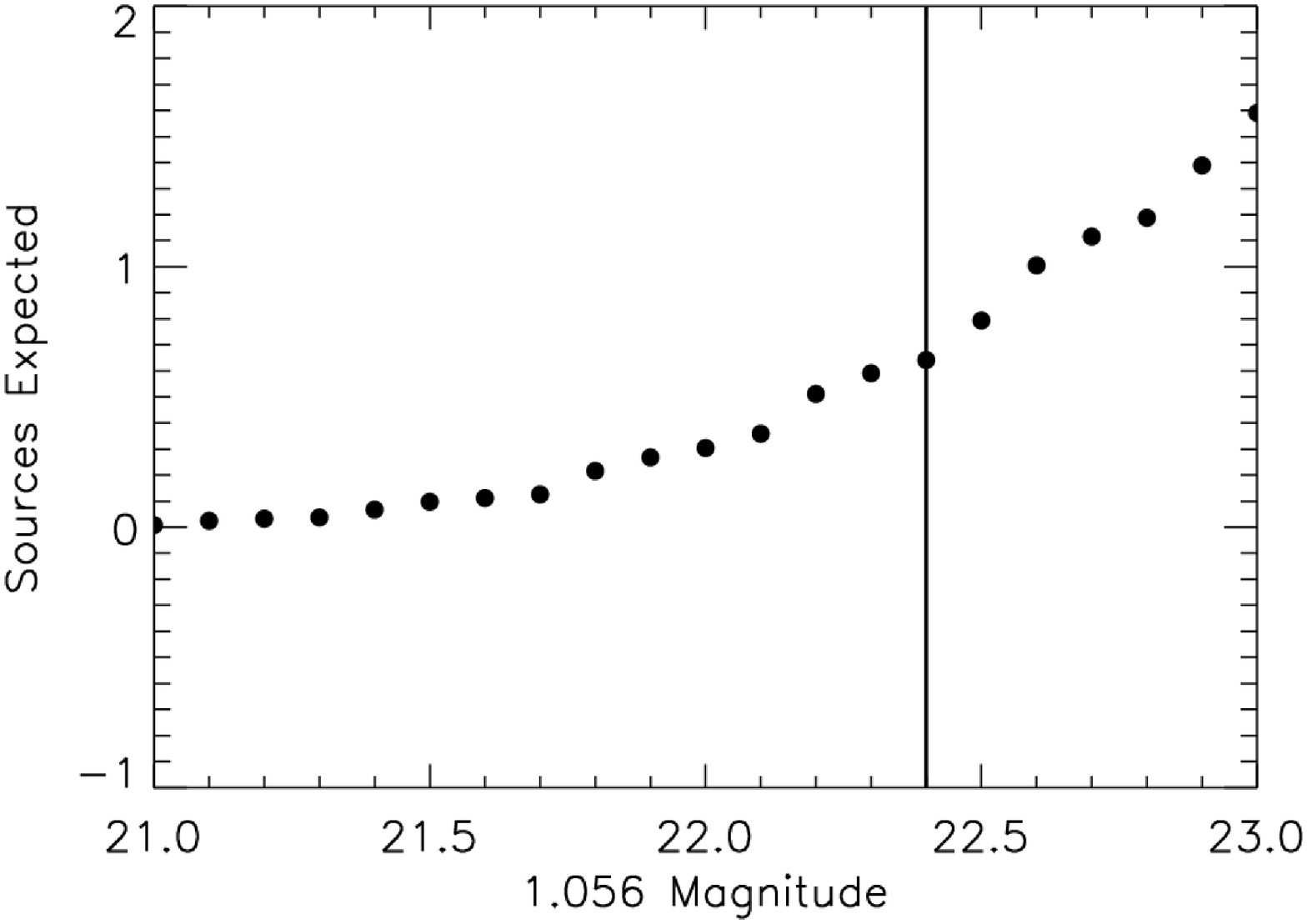}{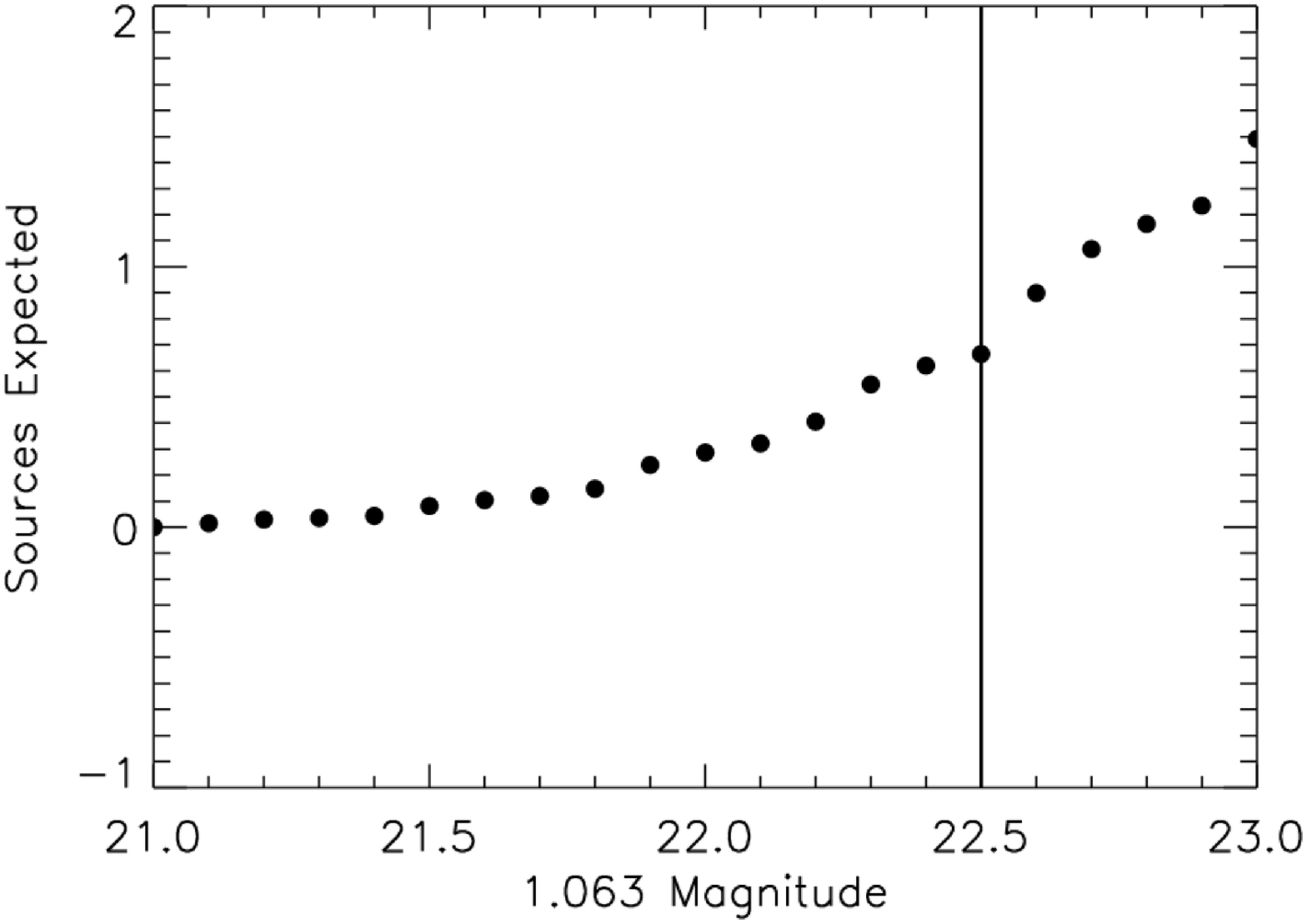}
\caption{Average results of the ten Monte Carlo simulation iterations,
 showing the number of \lya emitter sources we should expect at each
 magnitude bin.  Integrating out to our limiting magnitudes for each field
 (Section \ref{limmag}), indicated by the vertical line, we expect roughly 1 source per filter to
 be a real $z$ $=$ 7.7 \lya emitter if there is no evolution in the
 luminosity function between $z$ $=$ 6.5 and $z$ $=$ 7.7.}
\label{mcplots}
\end{center}
\end{figure}

This Monte Carlo simulation was run for each individual NEWFIRM UNB
filter and then iterated 10 times for each filter.  We then averaged
the results of those ten iterations, and the expected number of
sources per filter at each magnitude bin is shown in Figure
\ref{mcplots}.  Looking at the limiting magnitudes for each filter and
integrating out to that point, we
find that we expect $\sim$1 source per filter to be a true $z$ $=$
7.7 \lya emitter.  This result should be viewed with some caution,
however, as we have assumed that the \lya luminosity function does not
undergo significant evolution between redshifts of 6.5 and 7.7, and
that the emission line profiles of all \lya emitters at $z$ $=$ 7.7 are the
same as that of a \lya emission line profile at $z$ $=$ 5.7.  Moreover, this simulation makes use of the
Kashikawa et al. (2006) luminosity function, which is based upon detections
in only one field, the Subaru Deep Field.  We expect that there
will be field-to-field variations among \lya emitters, and thus the expected number of sources may
be different for the COSMOS field.  Various methods can be taken to
estimate the cosmic variance between fields of these \lya emitters;
for large survey volumes ($\sim$2 $\times$ 10$^5$ Mpc$^3$),
Tilvi et al. (2009) estimated this variance to be $\gtrsim$30\%.  We can
also utilize the method of Trenti \& Stiavelli (2008) to calculate the cosmic
variance expected assuming 2 intrinsic \lya sources as reported by our
Monte Carlo simulations.  Using the redshift interval of 0.1 for our
two filters and a Press-Schechter bias, the Trenti \& Stiavelli cosmic
variance
calculator\footnote{\texttt{http://casa.colorado.edu/$\sim$trenti/CosmicVariance.html}}
returns an expected cosmic variance of $\sim$58\% (this number drops
by $\sim$10\% with Sheth-Tormen bias).  In addition to cosmic
  variance, we note here that the most recent $z$=5.7 and 6.5 \lya LF
  results \citep{hu10,ouchi10,kash11} indicate that the predictions for
  the number densities of $z$=7.7 emitters based on the results of
  Kashikawa et al. (2006) may be up to a factor of 3 too large.  The uncertainty
  that remains in the lower-$z$ \lya LFs, taken in concert with the significant
  role that cosmic variance can play, requires the Monte Carlo
  simulations to be viewed with caution rather than interpreted at
  face value.

\section{$z$ $=$ 7.7 \lya LUMINOSITY FUNCTION} \label{lylf}

\begin{figure}
\begin{center}
\plotone{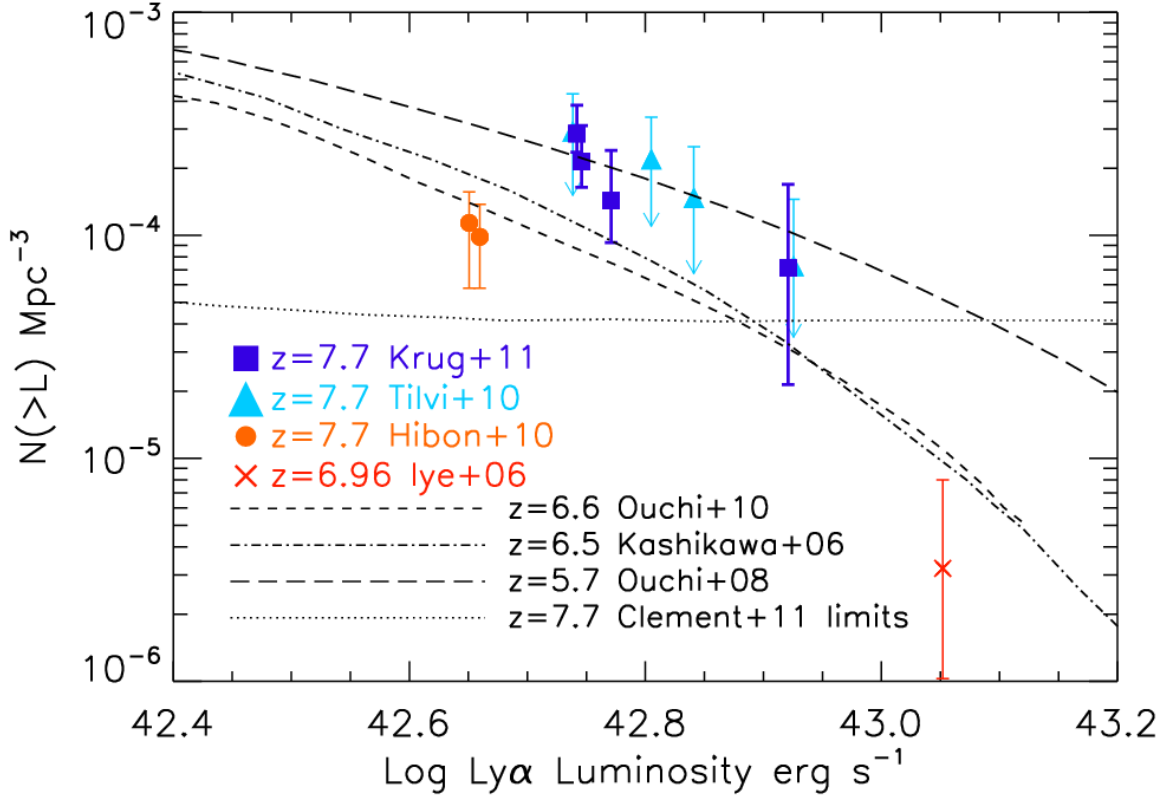}
\caption{\lya luminosity function compiled from data from this work
  (dark blue squares) and those previously published.  Errors are
  calculated using Poissonian statistics with Bayesian likelihood.
  The light blue
  triangles \citep{tilvi10} and orange circles \citep{hibon10}
  represent candidates at $z$=7.7 that have not been
  spectroscopically confirmed.  Recent follow-up spectroscopy by Cl\'ement et
  al. (2011) did not detect \lya
  emission from the five most luminous candidates of Hibon et
  al. (2010), so
  we have excluded these objects from the present figure.  The black dotted curve
  represents upper limits on the $z$=7.7 luminosity function from Cl\'ement et al. (2011).  The red X \citep{iye06} represents a
  spectroscopically confirmed $z$=6.96 \lya emitter.  The three black
  dashed and dashed-dotted curves
  represent best fit luminosity functions to spectroscopically
  confirmed \lya emitters at redshifts $z$=5.7~-~6.6.  Note that our
  \lya luminosity function agrees well with that of Tilvi et
  al. (2011).  Our two most luminous
  candidates are consistent within the error with either the $z$=5.7
  or $z$=6.5 luminosity function, and thus we do not see any evidence for
  significant evolution of the \lya luminosity function or neutral
  hydrogen fraction of the IGM between $z$=5.7
  and $z$=7.7.}
\label{lumfunc}
\end{center}
\end{figure}

The primary goal of this work is to constrain the observed luminosity
function (LF) of \lya emitters at $z$ $=$ 7.7.  As described in the
Introduction, our survey has the
advantage of being both wide and deep, leading to our volume and
limiting flux being better than or comparable to most other surveys at
this redshift or higher (Table \ref{othersurveys}).  To construct our
$z$ $=$ 7.7 \lya LF, we have used our four targets, which have
\lya line fluxes of 12.1, 8.6, 8.1, and 8.0 $\times$ 10$^{-18}$ erg
s$^{-1}$ cm$^{-2}$, all higher than our survey's limiting flux of 8 $\times$
10$^{-18}$ erg s$^{-1}$ cm$^{-2}$.  Using our survey volume of
$\sim$1.4 $\times$ 10$^4$ Mpc$^{3}$ per filter, we have constructed a cumulative
\lya LF which is shown in Figure \ref{lumfunc}.
In order to calculate the errors, we have used Poissonian statistics
with Bayesian likelihood (where the likelihood is equivalent to the probability), assuming each target is in a separate
luminosity bin.  The assumption in this error model is that the
luminosity distribution is uncorrelated - one galaxy having a certain
luminosity has no effect on the luminosity of another - and thus it
is safe to divide our luminosity bins this finely.  As we have a prior
probability distribution that is uniform in the expected number m (from Poisson), then the
lower and upper limits on our errors are both finite.  We have plotted our LF
along with the results of several other major surveys.  Two of these,
Tilvi et al. (2010) and Hibon et al. (2010), focus on the same redshift as
our work but have made use of alternate fields (CETUS and CFHT-LS D1,
respectively) and thus cosmic variance may come into play.  Neither
survey has published spectroscopic confirmation at this
time, although Cl\'ement et al. (2011) note that VLT spectroscopy failed
to detect \lya emission from the five most luminous Hibon et al. (2010)
candidates, a result which will be published soon; we plot the
Cl\'ement et al. (2011) upper limits for $z=7.7$ \lya LFs as well.  We also plot the spectroscopically confirmed Iye et al. (2006)
$z$ $=$ 6.96 \lya emitter.  In addition to individual data points, we
also plot three curves from previous, slightly lower redshift surveys,
which are based on best-fit functions to spectroscopically confirmed
data.  These include the Ouchi et al. (2008) $z$ $=$ 5.7 LF, as well as
the well-cited Kashikawa et al. (2006) $z$ $=$ 6.5 LF data.  We have also
included a new data set, the $z$ $=$ 6.6 LF from Ouchi et al. (2010),
which samples a different field from the Kashikawa $z$ $=$ 6.5 LF but
agrees well with that data set.  

Our data points agree quite well with the findings of
Tilvi et al. (2010), despite our surveys probing different fields,
but only the most luminous target matches with the Cl\'ement et al. (2011) upper limits
  for $z=$ 7.7 within the error; our data points also may seem inconsistent with recent Lyman
  break surveys (e.g., \citealt{pent11,schen11}).  Assuming that all four of our data points are real \lya emitters at
$z$ $=$ 7.7, we find good agreement with the $z$ $=$ 5.7 Ouchi et
al. (2008) function, yet our fainter luminosity objects do not agree
with the $z$ $=$ 6.5 LF \citep{kash06} nor the $z$ $=$ 6.6 LF
\citep{ouchi10}.  This could imply moderate evolution between
$z$ $=$ 7.7 and $z$ $=$ 6.6, yet it is interesting that in such a
case, the LFs at $z$ $=$ 7.7 and $z$ $=$ 5.7 would be in agreement.
If we instead take the more conservative approach indicated by our Monte Carlo simulation
results and assume only the two most luminous of our targets to be real \lya
emitters, then our results are consistent within error with both the
$z$ $=$ 5.7 Ouchi et al. (2008) function and the $z$ $=$ 6.5
Kashikawa et al. (2006) function.  Given that it is more probable statistically
that only these two candidates are real, we must take this case as the
more likely outcome until we have the ability to spectroscopically
verify our sources.  Thus, given the agreement of our two most luminous
emitters with the LFs at both lower redshifts, there is no conclusive
evidence for evolution of the \lya LF over the redshift range 5.7 $<$ $z$ $<$ 7.7.  

\section{SUMMARY} \label{conc}

We have utilized two custom-made UNB filters, at wavelengths 1.056 and
1.063 $\mu$m (FWHM 7.4 and 8.1 \AA, respectively) on the NEWFIRM
camera at the KPNO 4m Mayall telescope to perform a deep and wide
search for $z$ $=$ 7.7 \lya emitters in the COSMOS field.  Our study
comprised a co-moving volume of 2.8 $\times$ 10$^4$ Mpc$^{3}$ (survey
area $\sim$760 arcmin$^2$) and
probed down to a limiting flux of $\sim$8 $\times$ 10$^{-18}$ erg
s$^{-1}$ cm$^{-2}$ (50\% completeness limit), which is comparable to
or better than previous \lya searches at similar redshifts.

We used a very detailed selection procedure, making use of five
different quantitative parameters as well as qualitative methods to
ensure narrow-line detection and elimination of contaminants
(including 3 low-$z$ interlopers already known in the COSMOS catalog).
We were left with a total of four candidates (three detected in the 1.056
$\mu$m filter, one in the 1.063 $\mu$m filter), each detected at
5 sigmas or higher, down to line fluxes of 8
$\times$ 10$^{-18}$ erg s$^{-1}$ cm$^{-2}$.  Detailed Monte Carlo
simulations would suggest that up to two of these candidates are real.
If we assume only two real candidates, comparison of the resultant
\lya LF to the $z$ $=$ 5.7 Ouchi et al. (2008) and $z$ $=$ 6.5
Kashikawa et al. (2006) \lya LFs would indicate that there has been no
significant evolution of the \lya LF between 5.7 $<$ $z$ $<$ 7.7.
This result is consistent with the findings of Tilvi et al. (2010) -- 
that work used the same instrument, equivalent field of view, reached
similar flux limits, but was half the volume of our work,
owing to our extra filter.

To pin down the neutral hydrogen fraction at $z$ $=$ 7.7 and thus the
stage of the reionization process at that epoch, we will need more
detailed \lya LFs in order to accurately determine the characteristic
luminosity, $L^\star$, of these objects.  This will require a)
spectroscopic confirmations of these candidate high-redshift \lya
emitters, and b) surveys encompassing more fields,
as cosmic variance is likely to affect the number of emitters found in
each region.  The success of previous narrowband surveys at
identifying \lya emitters at lower redshift and the robustness of our
current set of candidates mean that the future of such high-$z$
studies is quite promising.  It is also encouraging to note that our
results match well with surveys in different fields.  The pursuit of
surveys such as these, along with the advances that should be brought
to the field by JWST and other new instruments, should provide a
bright future for the study of the Dark Ages.\\

H.B.K. and S.V. were supported by the NSF through contracts AST
0606932 and 1009583.
H.B.K. would like to acknowledge M. Coleman Miller for helpful
discussion of statistics and Bayesian error calculation, and is very
grateful to Peter
Capak for assistance with the COSMOS data catalog and discussion of
COSMOS PSFs and aperture correction methods.  Monte Carlo simulations
were performed using the University of Maryland's Yorp computing cluster.  The
observations reported here were obtained at the Kitt Peak National
Observatory, National Optical Astronomy Observatory, which is operated
by the Association of Universities for Research in Astronomy (AURA),
Inc., under cooperative agreement with the National Science Foundation.\\

%%%%%%%%%%%%%%%%%%%%%%%%%%%%%%%%%%%%%%%%%%%%%%%%%%%%%%%%%%%%%%%%%%%%%%%%%%%%%%%%%%%%

%%%%%%%%%%%%%%
% REFERENCES %
%%%%%%%%%%%%%%

%%%%%%%%%%%%%%%%%%%%%%%%%%%%%%%%%%%%%%%%%%%%%%%%%%%%%%%%%%%%%%%%%%%%%%%%%%%%%%%%%%%%

%%%%%%%%%%
% TABLES %
%%%%%%%%%%

\clearpage

\begin{deluxetable}{ccccc}
\tabletypesize{\footnotesize}
\tablecaption{Comparison to previous high-$z$ \lya
  searches\label{othersurveys}}
\tablewidth{0pt}
\tablehead{
\colhead{$z$} & \colhead{Survey Vol. (Mpc$^3$)} &
\colhead{Detec. Limits (erg s$^{-1}$ cm$^{-2}$)} & \colhead{\# of LAE
  Detec.} & \colhead{Refs} \\
\colhead{(1)} & \colhead{(2)} & \colhead{(3)} & \colhead{(4)} & \colhead{(5)}
}
\startdata
7.7 & 6.3$\times 10^4$ & 8.3$\times$10$^{-18}$ & 7 & Hibon+~2010\\
7.7 & 1.4$\times 10^4$ & 7$\times$10$^{-18}$ & 4 & Tilvi+~2010\\
7.7 & 2.8$\times 10^4$ & 8$\times$10$^{-18}$ & 4 & This work\\
%8.8 & 3 arcmin$^2$ & $\sim$10$^{-18}$ & 0 & Parkes+1994\\
8-10 & 35 & 2$\times$10$^{-17}$ & 6 & Stark+2007\\
8.8 & 6.3$\times 10^4$ & 1.3$\times$10$^{-17}$ & 0 & Cuby+2007\\
8.96 & 1.12$\times 10^6$ & 6$\times$10$^{-17}$ & 0 & Sobral+2009\\
\enddata
\tablecomments{Col.(1): \lya redshift probed.  Col.(2): Survey
  volume.  The volume of our study is 2x deeper than Tilvi et al. 2010 
  owing to our use of a second filter for candidate selection (each
  filter probes a volume
  of 1.4$\times 10^4$ Mpc$^3$).  Col.(3): Survey flux detection limit.
Col.(4): Number of candidate \lya emitters detected.  Col.(5): Survey reference. }
\end{deluxetable}

\begin{deluxetable}{ccccccc}
\tabletypesize{\footnotesize}
\tablecaption{Properties of four candidate \lya emitters. \label{cands_high}}
\tablewidth{0pt}
\tablehead{
\colhead{Cand. \#} & \colhead{RA} & \colhead{Dec} &
\colhead{Mag$_{\mathrm{AB}}$} & \colhead{Line Flux (erg s$^{-1}$
  cm$^{-2}$)} & \colhead{L$_\lya$ (erg s$^{-1}$)} & \colhead{EW$_\lya$
    (\AA)}\\
\colhead{(1)} & \colhead{(2)} & \colhead{(3)} & \colhead{(4)} &
\colhead{(5)} & \colhead{(6)} & \colhead{(7)}
}
\startdata
1 & 10:00:46.94 & +02:08:48.84 & 21.87 & 1.21$\times$10$^{-17}$ &
8.34$\times$10$^{42}$ & 7.32\\
2 & 10:00:20.52 & +02:18:50.04 & 22.25 & 8.55$\times$10$^{-18}$ &
5.90$\times$10$^{42}$ & 5.17\\
3$^{(a)}$ & 09:59:56.21 & +02:10:09.84 & 22.31 & 8.09$\times$10$^{-18}$ &
5.58$\times$10$^{42}$ & 4.89\\
4 & 10:00:48.79 & +02:09:21.24 & 22.39 & 8.00$\times$10$^{-18}$ &
5.52$\times$10$^{42}$ & 4.84\\
\enddata
\tablecomments{(a) Candidate \#3 fails the optical selection
  criterion following an aperture correction of the Subaru data; it is
  thus considered more uncertain than the others.  Col.(2): RA in J2000.  Col.(3): Dec in J2000.  Col.(4):
  AB magnitude, calculated using isophotal flux in SExtractor.
  Col.(5): Line flux, in the 1.056 $\mu$m band for candidates 1-3 and
  in the 1.063 $\mu$m band for candidate 4.  Col.(6): \lya luminosity,
  calculated assuming a redshift of $z$=7.7.  Col.(7): Lower limit rest-frame
  equivalent width of \lya emission, calculated as described in Section \ref{eqw}  equivalent widths.}
\end{deluxetable}

%%%%%%%%%%%%%%%%%%%%%%%%%%%%%%%%%%%%%%%%%%%%%%%%%%%%%%%%%%%%%%%%%%%%%%%%%%%%%%%%%%%%%

%%%%%%%%%%%
% FIGURES %
%%%%%%%%%%%

\end{document}